\documentclass[twocolumn,showpacs,epsfig,pra]{revtex4}

\usepackage{graphicx}
\usepackage{amssymb}
\usepackage{color}

\newcommand{\figsize}{0.45}

\begin{document}

\draft
\title{Exceptional points in coupled dissipative dynamical systems}
\author{Jung-Wan Ryu,$^1$ Woo-Sik Son,$^2$ Dong-Uk Hwang,$^2$ Soo-Young Lee,$^1$ and Sang Wook Kim$^3$}
\email{swkim0412@pusan.ac.kr}
\affiliation{$^1$School of Electronics Engineering, Kyungpook National University, Daegu 702-701, Korea\\
$^2$National Institute for Mathematical Sciences, Daejeon 305-811, South Korea\\
$^3$Department of Physics Education, Pusan National University, Busan 609-735, South Korea
}
\date{\today}

\begin{abstract}
We study the transient behavior in coupled dissipative dynamical systems based on the linear analysis around the steady state.
We find that the transient time is minimized at a specific set of system parameters and show that 
at this parameter set, two eigenvalues and two eigenvectors of Jacobian matrix coalesce at the same time,
this degenerate point is called the exceptional point.
For the case of coupled limit cycle oscillators, we investigate the transient behavior into the amplitude death state, and clarify 
that the exceptional point is associated with a critical point of frequency locking, as well as the transition of the envelope oscillation.
\end{abstract}
\pacs{05.45.Xt, 02.10.Ud}
\maketitle
\narrowtext

\section{Introduction}

In the eigenvalue problem of a non-Hermitian matrix, an exceptional point (EP) is a square-root branch point on a two-dimensional parameter space,
at which not only eigenvalues but also the associated eigenvectors coalesce \cite{Kat66, Hei12}.
The peculiar feature related to the EP is the exchange of eigenvalues and eigenvectors after a parameter variation encircling the EP once,
of which topological structure is same as that of M{\"o}bius strip \cite{Hei99}.
The EPs and relating interesting phenomena have mainly been studied in open quantum systems described by non-Hermitian Hamiltonians
such as atomic spectra in fields \cite{Lat95,Car07}, microwave cavity experiments \cite{Dem01,Dem03}, chaotic optical microcavities \cite{Lee09},
{\it PT}-symmetric quantum systems \cite{Ben98,Kla08,Rue10}, and so on.
Besides the open quantum systems, the EPs are also observed in coupled driven damped oscillators realized by electric circuits, which are
purely classical systems \cite{Hei04,Ste04}.

The amplitude death (AD) is the complete suppression of oscillations of the entire system when the nonlinear dynamical systems are coupled \cite{Sax12}.
The AD has been observed in many coupled dynamical systems and the AD is achieved by 
various types of coupling interaction, i.e., the diffusive coupling in mismatched oscillators \cite{Eli84,Mir90,Erm90,Aro90},
delayed coupling \cite{Red98,Red99,Red00a,Red00b,Zou13}, conjugate coupling \cite{Kar10}, dynamical coupling \cite{Kon03},
nonlinear coupling \cite{Pra03,Pra10}, etc.
The AD has also been studied in networks of coupled oscillators \cite{Erm90,Ata90} and variety topologies
such as a ring \cite{Dod04,Kon04}, small world \cite{Hou03}, and scale free networks \cite{Liu09}.
Recently, the suppressions of oscillations are strictly classified into
amplitude death and oscillation death, where the asymptotic steady state is homogeneous and inhomogeneous, respectively. \cite{Sax12, Kos13}

In this paper, we study the transient behaviors of coupled dissipative dynamical systems based on the linear analysis around the
steady state. We find that the systems show the largest damping rate at an EP, which comes from the intrinsic feature of a square-root branch point. 
For the case of coupled limit cycle oscillators, the transient behavior into the amplitude death state is studied.
We demonstrate that the EP is associated with a critical point of frequency locking, as well as the transition of the envelope oscillation.

This paper is organized as follows.
In Sec. II, we show the occurrence of EP in coupled damped oscillators and discuss the damping behavior around the EP in a pedagogical way.
In Sec. III, we present the transient behavior into the AD in coupled limit cycle oscillators,
and it is explained based on the existence of an EP.
Finally, we summarize our results in Sec. IV.

\section{Exceptional point in coupled damped oscillators}

\begin{figure}
\begin{center}
\includegraphics[width=\figsize\textwidth]{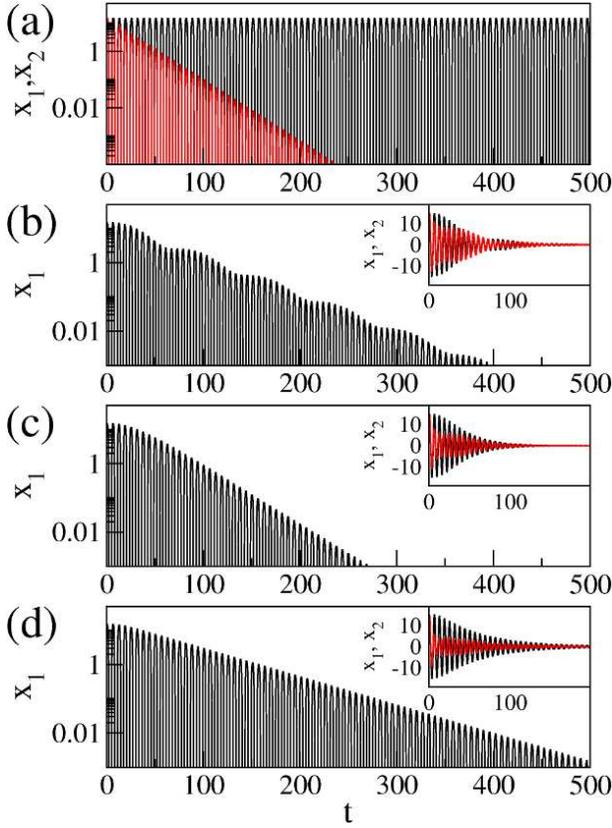}
\caption{(color online) Time series of $x_1$ (Black) and $x_2$ (Red) when $\omega_1 = \omega_2 =1.0$ and $\gamma_1 = 0.0$.
(a) No coupling case of $k=0.0$ with $\gamma_2 = 0.1$.
Coupling cases of $k=0.1$ with (b) $\gamma_2 = 0.1$, (c) $\gamma_2 = 0.2$, and (d) $\gamma_2 = 0.3$.
Insets show linearly scaled time series.}
\label{fig1}
\end{center}
\end{figure}

We consider the coupled damped oscillators,
\begin{eqnarray}
\label{cdo}
\nonumber
\ddot{x}_1+\gamma_1 \dot{x}_1+{\omega}_1^2 x_1 &=& -k x_2 , \\
\ddot{x}_2+\gamma_2 \dot{x}_2+{\omega}_2^2 x_2 &=& -k x_1 ,
\end{eqnarray}
where $\gamma_i$ and $\omega_i$ $(i=1,2)$ are damping ratio and undamped angular frequency of the $i$-th oscillator, and $k$ is the coupling constant.
Figure~\ref{fig1} shows the time series of $x_1$ and $x_2$ of Eq.~(\ref{cdo}) in the logarithmic scale when $\omega_1 =\omega_2 =1.0$ and $\gamma_1 =0$.
First, we consider uncoupled case, $k=0$. As we set $\gamma_1 =0$ and $\gamma_2 =0.1$, 
the time series of $x_1$ exhibits a stationary oscillation without damping, while
an exponential damping appears in the time series of $x_2$, as shown in Fig.~\ref{fig1}(a).
Next, we consider a finite coupling strength of $k=0.1$.
In Fig.~\ref{fig1}(b) with $\gamma_1 =0$ and $\gamma_2 =0.1$, both time series of $x_1$ and $x_2$ exhibit decays with envelope oscillations. Their decay rates, given by the slope of time series of $x_1$ and $x_2$ in the logarithmic plot, are equal.
As $\gamma_2$ increases from 0.1, the period of the envelope oscillation and the decay rate increase.
At $\gamma_2 \sim 0.2$, the envelope oscillation disappears and the decay rate reaches a maximum (see Fig.~\ref{fig1}(c)).
When $\gamma_2$ increases further, the decay rate decreases again. 
For example, the time series of the case with $\gamma_2 = 0.3$ is shown in  Fig.~\ref{fig1} (d).
Although the amplitude of two oscillators are different, as shown in the inset, 
their decay rates are equal. In our work, we concentrate on the case that each uncoupled oscillators have zero or weak damping ratio so that their dampings are underdamped.

In order to understand the variation of decay rate with $\gamma_2$
and its maximum at $\gamma_2 \sim 0.2$, 
we analyze the eigenvalues of a stability matrix around the origin.
Eq.~(\ref{cdo}) can be rewritten as
\begin{eqnarray}
\label{cdo2}
\nonumber
\dot{x}_1 &=& y_1 , \\\nonumber
\dot{y}_1 &=& -\gamma_1 y_1 -{\omega}_1^2 x_1 - k x_2 , \\\nonumber
\dot{x}_2 &=& y_2 , \\
\dot{y}_2 &=& -\gamma_2 y_2 -{\omega}_2^2 x_2 - k x_1 .
\end{eqnarray}
This set of equations is represented by a vector equation,
$\dot{\vec{z}}(t) = M{\vec{z}}(t)$, 
where $\vec{z}(t)=(x_1(t), y_1(t), x_2(t), y_2(t))^T$.
The stability matrix $M$ is then given by
\begin{equation}
\label{cdo_mat}
M = \left(
\begin{array}{cccc}
0 & 1 & 0 & 0 \\
-{\omega}_{1}^2 & -\gamma_1 & -k & 0 \\
0 & 0 & 0 & 1 \\
-k & 0 & -{\omega}_{2}^2 & -\gamma_2 \\
\end{array}
\right).
\end{equation}
The eigenvalues $\lambda_l$ of $M$ are complex numbers, because the matrix $M$ is  non-Hermitian.
Since the time evolution of an eigenvector $\hat{e}_l$ is given as $ e_l(t)=\hat{e}_l \exp (\lambda_l t)$, 
the real and imaginary parts of the eigenvalues correspond to the decay rates and the angular frequency of
the corresponding time series, respectively.

\begin{figure}
\begin{center}
\includegraphics[width=\figsize\textwidth]{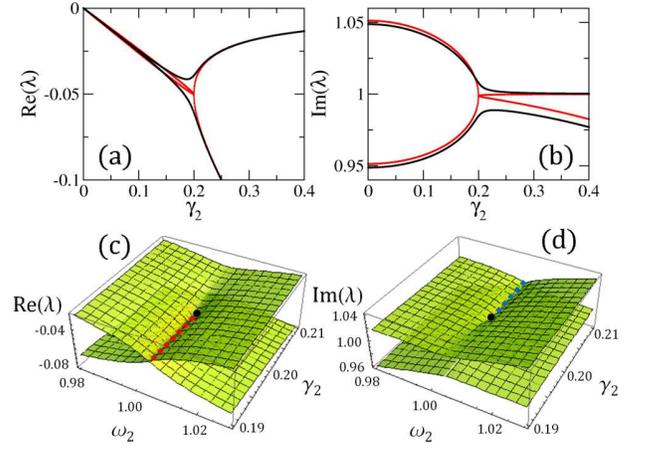}
\caption{(color online). (a) Real and (b) imaginary parts of two eigenvalues of which imaginary parts are positive
as a function of $\gamma_2$ when $\omega_2 = 1.0$ (black) and $\omega_2 = 1.005$ (red) with $\gamma_1 =0.0$ and $\omega_1 = 1.0$.
(c) Real and (d) imaginary parts of the eigenvalues near EP as functions of $\omega_2$ and $\gamma_2$
when $\omega_1 = 1.0$ and $\gamma_1 = 0.0$.
The black circle, red dotted line, and blue dotted line represent the EP, real value crossing line, and imaginary value crossing line,
respectively.}
\label{fig2}
\end{center}
\end{figure}

The complex eigenvalues with positive imaginary parts are shown as a function of $\gamma_2$ in Fig.~\ref{fig2}(a) and (b).
When $\gamma_2 < 0.2$, real parts of two eigenvalues are very close but their imaginary parts are quite different,
this means that the dynamics of eigenvectors would show almost same decay rate and different angular frequencies.
In this range, the time series of $x_1$ and $x_2$ would show a constant overall slope given by the close real parts,
but they would have an oscillatory envelope whose frequency is determined by the difference of the imaginary parts
of eigenvalues. This behavior has been shown in Fig.~\ref{fig1}(b). 
As $\gamma_2$ approaches to $0.2$, the real parts of two eigenvalues decrease and the imaginary parts become closer with each other,
which corresponds to the time series with a faster decay and a longer period of envelope oscillation, respectively.

As $\gamma_2$ goes further beyond $0.2$, two real parts start to split but the difference of two imaginary parts become small.
The splitting of two real parts indicates that the time series can be characterized by
a combination of fast and slow decays. The fast decay might be seen only in the short time behavior
and the slow decay, corresponding to the larger real part, dominates the long time behavior of the time series.
Thus, although two imaginary parts are still different, there is no envelope oscillation due to the fast suppression
of one eigen-component with the lower real part (see Fig.~\ref{fig1}(d)).
Note that the larger real part, governing long-time behavior, has a minimum value around at $\gamma_2 \sim 0.2$, which
explains the maximum decay rate observed in Fig.~\ref{fig1}(c).

Note that two complex eigenvalues are very close at $\gamma_2 \sim 0.2$ as shown by the black lines
in Fig.~\ref{fig2} (a) and (b). We can expect that there should be a degenerate point, called exceptional point (EP) \cite{Kat66, Hei12},
where two complex eigenvalues coalesce, in the system parameter space. 
By adjusting $\omega_2$ a bit as $\omega_2 = 1.005$, we find an EP at $(\omega_2, \gamma_2 ) \sim (1.005, 0.2)$, which
is shown by the red lines in Fig.~\ref{fig2} (a) and (b).
It is well known that two eigenvectors also coalesce at the EP and mathematically the EP is the square-root branch point.
The EP can be characterized by a peculiar eigenvalue surfaces in a parameter plane.
In Fig.~\ref{fig2} (c) and (d), the surfaces of the two eigenvalues are plotted in $(\omega_2 ,\gamma_2)$ plane.
Topology of the surface explains the exchange of two eigenvalues
for a parameter variation encircling the EP \cite{Hei99}. It is emphasized that the larger real part becomes a local minimum
at the EP, indicating the local maximum decay rate in the parameter plane.

\section{Exceptional point and amplitude death in coupled limit cycle oscillators}

In this section, we study the role of the EP when the amplitude death (AD) occurs in coupled limit cycle oscillators.
Let us start with the following system of two Stuart-Landau limit-cycle oscillators with diffusive coupling:
\begin{eqnarray}
\label{clco}
\nonumber
\dot{z}_1=(R_1 + i \omega_1 - {|z_1|}^2 ) z_1 + k (z_2 - z_1), \\
\dot{z}_2=(R_2 + i \omega_2 - {|z_2|}^2 ) z_2 + k (z_1 - z_2),
\end{eqnarray}
where $z_j$ are complex variables, $\omega_j$ are the intrinsic angular frequencies of uncoupled $j$-th limit cycle oscillators,
and $k$ is the coupling strength. Without coupling ($k=0$), two limit cycle oscillators are attracted to the
limit cycle with radii $\sqrt{R_j}$ for $R_j > 0$ and the origin for $R_j < 0$.
Stuart-Landau limit-cycle oscillator is renowned as a paradigmatic model for studying the AD in coupled nonlinear oscillators because it is a prototypical system exhibiting a Hopf bifurcation
that can reveal universal features of many practical systems.
For instance, a variety of spatio-temporal periodic patterns can be created in two-dimensional lattice of delay-coupled Stuart-Landau oscillators \cite{Kan14}.

\subsection{The amplitude death in coupled limit cycle oscillators}

It has been well known that the AD occurs in coupled limit cycle oscillators at proper $k$
if the $\Delta \omega=\omega_2 - \omega_1$ is sufficiently large when $R_1 = R_2 = 1.0$ \cite{Mir90,Erm90,Aro90}.
In order to obtain the AD region in the parameter space ($\Delta \omega, k$),
we calculate the Jacobian matrix $J$ at the origin, which is given by
\begin{equation}
\label{clco_mat}
J = \left(
\begin{array}{cccc}
R_1 - k & -\omega_1 & k & 0 \\
{\omega_{1}} & R_1 - k & 0 & k \\
k & 0 & R_2 - k & -\omega_2 \\
0 & k & {\omega_{2}} & R_2 - k \\
\end{array}
\right).
\end{equation}
The eigenvalues $\lambda$ of $J$ are complex numbers because the Jacobian matrix $J$ is a non-Hermitian matrix.
That is, the real and imaginary parts are the decay (or growing) rates and the angular frequency of the orbit
near the origin, respectively.

\begin{figure}
\begin{center}
\includegraphics[width=\figsize\textwidth]{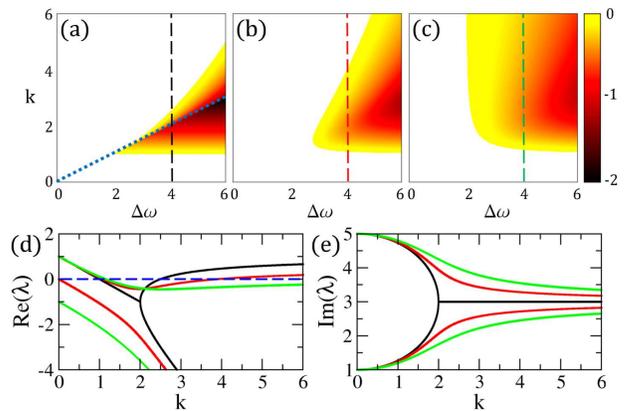}
\caption{(color online). Maximal values of real parts of eigenvalues with (a) $R_2 = 1.0$, (b) $R_2 = 0.0$, and (c) $R_2 = -1.0$
when $R_1 = 1.0$.
The colored and white region represent negative and positive values, respectively.
The blue dotted line represents the EP.
(d) Real and (e) imaginary parts of two eigenvalues of which imaginary parts are positive
as a function of $k$ when $\Delta \omega=4.0$ and $R_1 = 1.0$.
Black, red, and green curves represent the cases of $R_2 = 1.0$, $0.0$, and $-1.0$, respectively.
}
\label{fig3}
\end{center}
\end{figure}

The occurrence of AD is determined by the stability of the origin, which is related to the maximal value
of the real parts of complex eigenvalues.
If the maximal value is negative, the origin is stable fixed point and therefore the system exhibits the AD.
The colored region in Fig.~\ref{fig3}(a)-(c) where the maximal value is negative represent the AD regions
when $R_2 =1.0$, $0.0$, and $-1.0$, respectively, with $R_1 = 1.0$.
As $R_2$ decreases from $1.0$ to $-1.0$, the AD region becomes larger.
Figure~\ref{fig3}(d) clearly shows the transition between positive and negative values of maximal real parts
as a function of $k$ when $\Delta \omega$ is fixed.

\subsection{The exceptional point in coupled limit cycle oscillators}

Similarly as the case of coupled damped oscillators, there also exists an EP in the coupled limit cycle oscillators. The EP occurs at $k=2.0$ when $\Delta \omega = 4.0$ and $R_1 =R_2 =1.0$,
which is the double root position in Fig.~\ref{fig3}(d) and (e).
Considering $R_1 =R_2 =R$, four eigenvalues of Eq.~(\ref{clco_mat}) are given by
\begin{eqnarray}
\label{clco_ev}
\nonumber
-k+R \pm \sqrt{ -\frac{(\Delta \omega)^2}{2} +k^2 - \triangle}, \\
-k+R \pm \sqrt{ -\frac{(\Delta \omega)^2}{2} +k^2 + \square},
\end{eqnarray}
where $\triangle = \frac{\Delta \omega}{2} (\sqrt{{\Delta \omega}^2 -4 k^2} + 2 \omega_{1})
+ \omega_1 (\sqrt{{\Delta \omega}^2 -4 k^2} + \omega_1 )$ and
$\square = \frac{\Delta \omega}{2} (\sqrt{{\Delta \omega}^2 -4 k^2} - 2 \omega_{1})
+ \omega_1 (\sqrt{{\Delta \omega}^2 -4 k^2} - \omega_1 )$, respectively.
From the condition for EP, i.e., $\triangle = - \square$, the analytic condition for the existence of EP is given by
\begin{eqnarray}
\label{ep}
R_1 = R_2 , ~ k= \Delta \omega /2.
\end{eqnarray}
The eigenvectors also coalesce at this condition.
According to the Eq.~(\ref{ep}),
the EP occurs on the line in the parameter space ($\Delta \omega, k$) when $\Delta R =R_2 - R_1 = 0.0$ as shown in Fig.~\ref{fig3}(a).
If $\Delta \omega$ is fixed,
it is expected that a system shows the fastest attracting to the AD state on the condition of EP, $k=\Delta \omega /2$.
Because the decaying rate to the AD state can be considered as a maximal value of Re($\lambda$)
and the maximal value of Re($\lambda$) has its minimum at the condition of EP, $k = \Delta \omega /2$ [cf. Fig.~\ref{fig3}(d)].
In addition, there is transition of transient behavior to the AD state on the EP, which is the transition
between decaying with envelope oscillation due to the effective beat note for $k< \Delta \omega /2$
and decaying without envelope oscillation for $k> \Delta \omega /2$.
It is noted that as $R_2$ decreases from $1.0$, the AD region becomes larger, while the fastest attracting to AD state
occurs on the EP when $R_2 = 1.0$.

\begin{figure}
\begin{center}
\includegraphics[width=\figsize\textwidth]{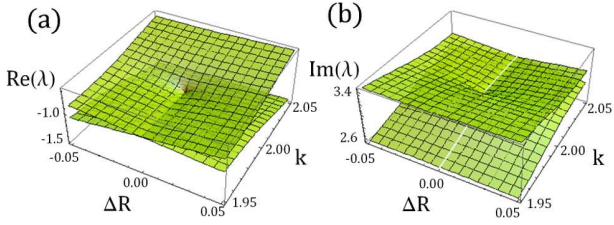}
\caption{(color online). (a) Real and (b) imaginary parts of two eigenvalues
near the EP at $(\Delta R, k)=(0.0, 2.0)$ when $\Delta \omega=4.0$}
\label{fig4}
\end{center}
\end{figure}

Figure~\ref{fig4} shows the complex eigenvalues near the EP in the parameter space ($\Delta R, k$), which is the singular point.
We note that there are the same topological structures of eigenvalues near the EPs in coupled limit cycle oscillators with the parameter
space ($\Delta R, \Delta \omega$) when $k = 2.0$ according to the Eq.~(\ref{ep}).

\subsection{Numerical results}

\begin{figure}
\begin{center}
\includegraphics[width=\figsize\textwidth]{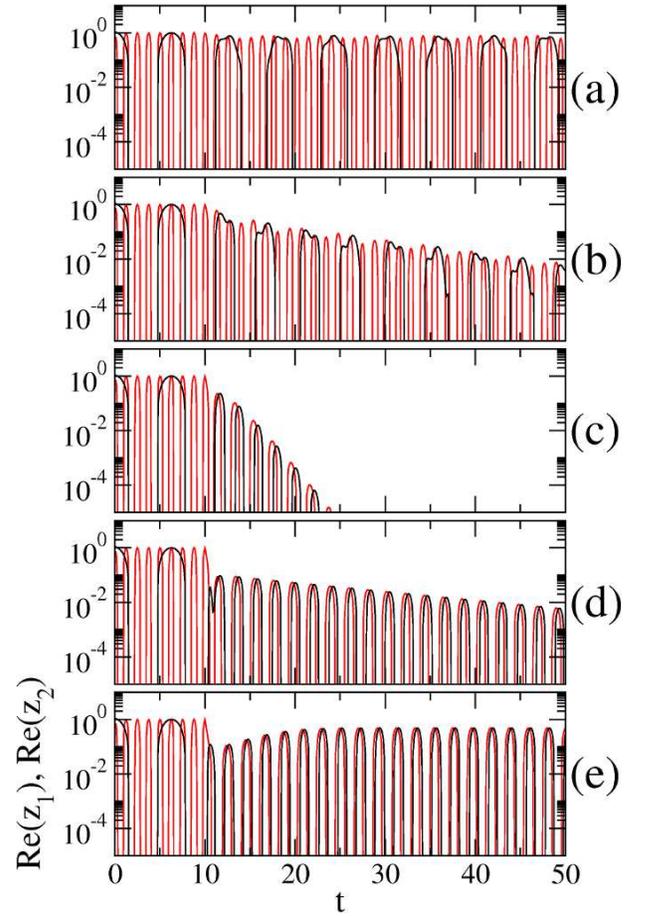}
\caption{(color online). Time series of real parts of $z_1$ (black) and $z_2$ (red)
with (a) $k=0.5$, (b) $1.1$, (c) $2.0$, (d) $2.4$, and (e) $3.0$ when $\Delta \omega=4.0$ and $R_1 = R_2 = 1.0$.}
\label{fig5}
\end{center}
\end{figure}

In order to confirm the role of the EP expected in the previous subsection, we obtain the time series
of $z_1$ and $z_2$ as $k$ increases.
Figure~\ref{fig5} shows the time series of real parts of $z_1$ and $z_2$ with different $k$
when $\Delta \omega =4.0$ and $R_1 =R_2 =1.0$.
The coupling is turned on at $t=10.0$, i.e., $k=0.0$ when $t<10.0$.
At $k=0.5$, neither AD nor $1:1$ frequency locking occurs because of small coupling strength.
Let us remind that the AD and 1:1 frequency locking occur,
when the maximal value of Re($\lambda$) in Fig.~\ref{fig3}(d) is lower than zero
and the pair of Im($\lambda$) in Fig.~\ref{fig3}(e) equal each other, respectively.
At $k=1.1$, the AD occurs with transient behavior of envelope oscillation but there is no frequency locking on the transient behavior.
At $k=2.0$, the AD occurs without envelope oscillatory transient behavior and the decay is fastest because this is the condition of the EP.
The $1:1$ frequency locking on the transient behavior is also shown.
At $k=2.4$, the AD occurs without envelope oscillation and there is frequency locking on the transient behavior.
The decay is slower than that in the case of $k=2.0$.
At $k=3.0$, the AD does not occur but there is frequency locking.
In the AD region ($1.0 < k < 2.5$), the EP is the transition point between decaying {\it with} and {\it without} envelope oscillations.
Also, in this region, the EP is the transition point for frequency locking.
The imaginary parts of eigenvalues relating to the frequencies change two different values into one value via the EP when $R_2 =1.0$.
If $R_1 \ne R_2$, two different frequencies are changed into two close frequencies not an identical frequency and therefore
there is no exact frequency locking of transient.

\begin{figure}
\begin{center}
\includegraphics[width=\figsize\textwidth]{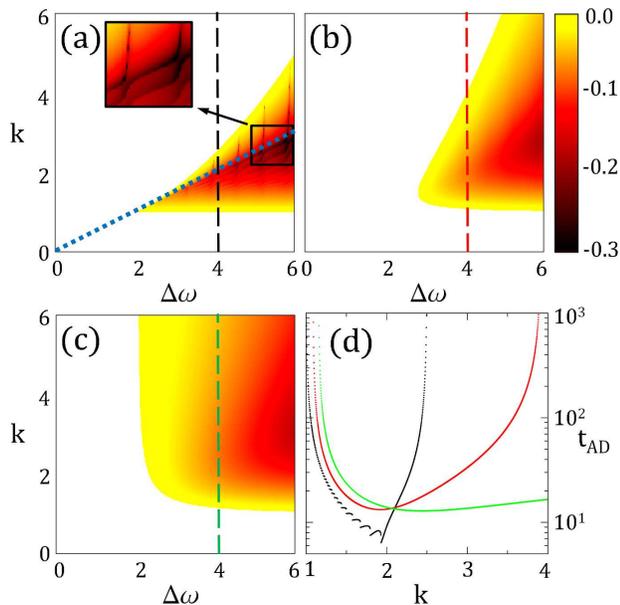}
\caption{(color online). $-(1/t_{AD})$ as functions of $\Delta \omega$ and $k$
with (a) $R_2 =1.0$, (b) $0.0$, and (c) $-1.0$ when $R_1 = 1.0$.
The colored and white region represent the AD and non-AD regions, respectively.
The blue dotted line represents the EP.
(d) $t_{AD}$ as a function of $k$ when $R_2 =1.0$ (black), $0.0$ (red), and $-1.0$ (green) when $\Delta \omega =4.0$.
}
\label{fig6}
\end{center}
\end{figure}

The important role of EP in AD is that the condition of EP guarantees the fastest attracting time to the origin, i.e., the AD state.
We investigate the attracting time to the AD state, denoted by $t_{AD}$. Here, $t_{AD}$ is calculated by followings;
If, at time $t$, the radii of two oscillators firstly become smaller than $c_{AD}$, i.e., the criterion for the AD state,
and continue to be smaller than $c_{AD}$ for $200$ seconds,
then $t_{AD}$ equals to $t-t_{on}$ where $t_{on}$ is the time when the coupling is turned on.
We set $c_{AD}=0.001$ and $t_{on}=10.0$.
Figure~\ref{fig6} (a)-(c) show $-(1/t_{AD})$, with various $R_2$ when $R_{1}=1.0$, on the parameter space $(\Delta \omega, k)$.
Figure~\ref{fig6} (d) shows $-(1/t_{AD})$ as a function of $k$ when $R_{1}=1.0$ and $\Delta \omega=4.0$ and the local minimum appears more clear when the parameters of system are closer to the EP.
Contrary to the expectation from the maximal real parts of eigenvalues in Fig.~\ref{fig3}, there are many wrinkled patterns when $R_2 = 1.0$.
The wrinkled patterns gradually disappear as $R_2$ decreases and then there is no patterns when $R_2 =-1.0$.
The different $t_{on}$ which means the different initial conditions makes the different wrinkled patterns.
The wrinkled patterns when $k<\Delta\omega/2$ are caused by the oscillatory transient behavior.
However, the reason of the wrinkled patterns when $k>\Delta\omega/2$ is that the transition from fast decay to slow decay occurs
when the amplitudes of the oscillators are smaller than our critical value $c_{AD}$
and therefore the patterns disappear if $c_{AD}$ is sufficiently small.

\begin{figure}
\begin{center}
\includegraphics[width=\figsize\textwidth]{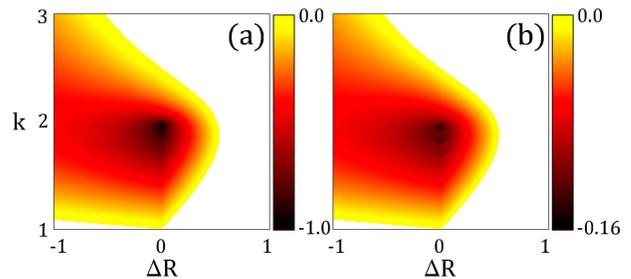}
\caption{(color online). 
(a) Maximum of real parts of eigenvalues and (b) $-(1/t_{AD})$ as functions of $\Delta R$ and $k$ with $\Delta \omega = 4.0$.
The colored and white region represent the AD and non-AD regions, respectively.
}
\label{fig7}
\end{center}
\end{figure}

Figure~\ref{fig7} shows the maximal values of real parts of eigenvalues and $-(1/t_{AD})$ with the parameter space ($\Delta R$, $k$)
when $\Delta \omega = 4.0$.
As shown in Fig.~\ref{fig4}, the EP exists at ($\Delta R$, $k$)=(0.0, 2.0),
where the maximal values of real parts of eigenvalues are local minimum as shown in Fig.~\ref{fig7} (a).
$t_{AD}$ are also local minimum at the EP.
In Fig.~\ref{fig7}(b), the wrinkled patterns exist at $\Delta R = 0.0$ but they disappear as $\Delta R$ deviates from $0.0$.
In principle, for the long time behavior, the oscillation behavior such as underdamped case exists only on the line,
$\Delta R=0$ and $k<\Delta \omega /2$,
because the real parts of two eigenvalues are same on the line in the parameter space ($\Delta R$, $k$).
Different real parts of two eigenvalues mean the system has two different decay rates and therefore only one frequency is dominant
for a long time behavior.
It is noted that the EP is not local minimal point on the parameter space ($\Delta \omega$, $k$) because the EP forms the lines
as shown in Fig.~\ref{fig3} (a) and Fig.~\ref{fig6} (a).
That is, the maximal values of real parts of eigenvalues decrease as the $\Delta \omega$ increases on the EP line, $k=\Delta \omega /2$. 

\section{Summary}

We have studied the exceptional point in dynamical systems and investigated the role of the exceptional point in the transient behaviors of amplitude death in coupled limit cycle oscillators.
The exceptional point is associated with a critical point of frequency locking as well as the transition of the envelope oscillation,
which also gives the fastest decay to the amplitude death in coupled limit cycle oscillators.
In addition, for other examples (two Van der Pol oscillators interacting through mean-field diffusive coupling, and coupled system of the R{\"o}ssler and a linear oscillator),
we have obtained the largest decay rates and transition behaviors at exceptional point (not shown here).
As a result, the transient behaviors related to the exceptional point appear commonly for the coupled dissipative dynamical systems, independent of the specific properties of systems. 
We expect the exceptional point is important to the study on the various disciplines such as
the nonequilibrium statistical mechanics \cite{Zwa01} and transient chaos \cite{Lai11,Mot13}
because the exceptional point is not related to the stationary states but the transient behaviors.

\section*{Acknowledgment}

This research was supported by Basic Science Research Program through the National Research Foundation of Korea (NRF) funded
by the Ministry of Education (No.2012R1A1A4A01013955 and No.2013R1A1A2011438).
This research was supported by National Institute for Mathematical Sciences (NIMS) funded by the Ministry of Science, ICT \& Future Planning
(A21501-3; B21501).

\end{document}